\documentclass[namedreferences]{solarphysics}
\usepackage[optionalrh]{spr-sola-addons}  
\usepackage{graphicx}                    
\usepackage{amssymb}                    
\usepackage{color}                       
\usepackage{url}                         

\begin{document}

\begin{article}

\begin{opening}

\title{On the Collective Magnetic Field Strength and Vector Structure of Dark Umbral Cores Measured by the Hinode Spectropolarimeter}

%
\author{T.A.~\surname{Schad}$^{1,2}$}

%
\runningauthor{Schad}
\runningtitle{On the Collective Vector Magnetic Field Structure of Dark Umbral Cores}

%

\institute{$^{1}$ 	Department of Planetary Sciences, 
				University of Arizona, Tucson, AZ 85721, USA	\\
				$^{2}$ 	National Solar Observatory (NSO), 
				950 N Cherry Ave., Tucson, AZ 85719, USA
				email: \url{schad@nso.edu}
	}

\begin{abstract}
We study 7530 sunspot umbrae and pores measured by the \textit{Hinode Spectropolarimeter} (SP) between November 2006 and November 2012.  We primarily seek confirmation of the long term secular decrease in the mean magnetic field strength of sunspot umbrae found by Penn and Livingston (2011, \textit{IAU Symp.} \textbf{273},126) between 1998 and 2011.  The excellent SP photometric properties and full vector magnetic field determinations from full-Stokes Milne-Eddington inversions are used to address the interrelated properties of the magnetic field strength and brightness temperature for all umbral cores.  We find non-linear relationships between magnetic field strength and umbral temperature (and continuum contrast), as well as between umbral radius and magnetic field strength.  Using disambiguated vector data, we find that the azimuths measured in the umbral cores reflect an organization weakly influenced by Joy's law.  The large selection of umbrae displays a log-normal size spectrum similar to earlier solar cycles.  Influenced by the amplitude of the solar cycle and the nonlinear relationship between umbral size and core magnetic field strength, the distribution of core magnetic field strengths, fit most effectively with a skew-normal distribution, shows a weak solar cycle dependence.  Yet, the mean magnetic field strength does not show a significant long term trend.
\end{abstract}


\end{opening}


\section{Introduction}\label{sec:intro}

Despite their often complex configurations and dynamic evolution, simplistic properties of sunspots when viewed in collection set important constraints on the activity cycles of our Sun, as well as other stars.  The record of the waxing and waning number of individual sunspots and sunspot groups forms one of the longest astrophysical studies to date.  Sunspots, of course, are not individual.  Subject to Gauss's law for magnetism, sunspots form within bipolar active regions that emerge from below the photosphere to establish paired groups of opposite polarity spots and/or plage.  This process is organized according to Hale's polarity law, Joy's law for the tilt of bipolar active regions, and an equator-ward drift of a northern and southern band of active latitudes (see, \textit{e.g.} \inlinecite{stix2004}).  Each of these solar cycle characteristics inform us about the mechanisms generating stellar activity; yet, much is unknown, including what leads to long-term declines in activity such as the Maunder Minimum between about 1645 to 1715 \cite{eddy1976}.

Recently, there has been renewed interest in the average magnetic and thermal properties of sunspots over the solar cycle.  Measurements from the \textit{McMath-Pierce Telescope} of the magnetic field strength and continuum brightness within the darkest portion of sunspot umbrae (\textit{i.e.} umbral cores) exhibit on average a long-term decline in field strength and a long-term increase in brightness between 1998 and 2011 \cite{penn2006,penn2011,livingston2012}.  These data examine the Zeeman-splitting of the Stokes-I $\sigma$ components of the Fe I normal Zeeman triplet at 1564.8 nm (Land\'{e} factor, g = 3).  In contrast, a study of archived magnetic field measurements gleaned from polarized states of visible spectral lines from the former USSR indicates that the maximum field strength in sunspots rise and fall during the solar cycle and do not show a long-term decline between 1957 and 2011 \cite{pevtsov2011}.  This cyclic behavior is supported by the line-of-sight (LOS) umbral magnetic flux measurements from \inlinecite{watson2011}.  The primary difference between these studies and that of \inlinecite{livingston2012} is that only the largest spots on any given observing day are included in the temporal averages of the umbral maximum magnetic field strength by \inlinecite{pevtsov2011} and \inlinecite{watson2011}, while \inlinecite{livingston2012} include all observed spots.

Determining whether the average thermal and magnetic properties of sunspots vary in time intricately depends upon the accurate characterization of the distributed magnetic field, thermal, and size parameters of sunspots.  The maximum umbral field strength and minimum intensity are closely related with the size of the sunspot umbra \cite{kopp1992,mathew2007,schad2010}, while the maximum magnetic field strength and minimum intensity of sunspot umbrae are well correlated with each other \cite{norton2004,schad2010}.  \inlinecite{nagovitsyn2012} suggest that, both the solar cycle variation of the maximum magnetic field of the largest sunspots during a solar cycle inferred by \inlinecite{pevtsov2011} and the long-term decline in the average sunspot magnetic field reported by \inlinecite{penn2006}, can be explained by temporal variations in the relative distribution of small and large sunspots, as seen in full disk synoptic images recorded at the Kislovodsk Mountain Astronomical Station of Pulkovo Observatory.  This is in disagreement with studies of the relative size distribution of umbrae measured between 1917 and 1982 using Mount Wilson whitelight images \cite{bogdan1988}, which exhibits a log-normal distribution that varies only in amplitude during the solar cycle. The invariant log-normal size distribution \cite{bogdan1988} is further supported by measurements from the Greenwich Observatory \cite{baumann2005} and the \textit{Kitt Peak Vacuum Telescope's Spectromagnetogragh} (KPVT/SPM) \cite{schad2010}.

\inlinecite{schad2010} and \inlinecite{mathew2007} discuss the role of the size dependence of the umbral intensity measurements to argue against a solar cycle dependence of the average umbral core intensity.  \citeauthor{schad2010} also show, using LOS magnetic flux measurements, that in an automated selection of sunspot umbrae between 1993 and 2004, the average magnetic field strength of sunspot umbrae does not exhibit any significant temporal changes, nor do the relationships between umbral core intensity, size, and maximum magnetic field strength.  The role of the size dependence is not addressed by \inlinecite{penn2006}, who used the \textit{McMath-Pierce Telescope} data set, as umbral sizes are not measured.  Although the authors assume that the measurements are not biased by selection effects, this is not explicitly demonstrated. 

A great disadvantage of the studies by \inlinecite{schad2010} and \inlinecite{watson2011} is, in contrast, the primary advantage of the \inlinecite{penn2006} data set.  The Fe I 1564.8 nm intensity spectra is sensitive to the \textit{total} magnetic field strength within the sunspot umbrae, since its Zeeman $\sigma$ components are well resolved in most umbrae.  Umbral magnetic field measurements from KPVT/SPM and the \textit{Michelson Doppler Imager} on board the \textit{Solar and Heliospheric Observatory} (SOHO/MDI) only measure the LOS magnetic flux, and are subject to other errors inherent to the intensity-difference and center-of-gravity magnetograph techniques.  Furthermore, to estimate the total magnetic field in umbral cores, one must make the assumption of a unity fill fraction and that the umbral magnetic field is vertical (\textit{i.e.} radial) within dark umbral cores, which until now has not been tested for a large collection of umbrae (see Section~\ref{sec:vector_distrib}).

Modern full-Stokes spectropolarimeters infer the true umbral magnetic field strength via full Stokes analysis of spectral lines.  Although no synoptic full-disk vector measurements are available for the full time period of Penn and Livingston's \citeyear{2006} measurements, the six years (November 2006 - present) of high resolution full-Stokes measurements from the \textit{Spectropolarimeter} (SP) \cite{tsuneta2008,lites2013,lites_ichimoto2013} onboard the \textit{Hinode} spacecraft \cite{kosugi2007} offers a unique chance to investigate the collective thermal and magnetic field vector properties of sunspot umbrae.  Below we discuss the full vector structure of dark umbral cores measured by \textit{Hinode}/SP, presenting in detail the interrelated magnetic field, thermal, and size parameters of sunspot umbrae (see Sections~\ref{sec:B_I_TEMP} and~\ref{sec:size_depend}). We find an agreement between our measured distributions of umbral sizes and the study of \inlinecite{bogdan1988}, and, thus, we argue that our selection adequately samples the solar cycle (see Section~\ref{sec:temporal_effects}).  The core magnetic field strengths are skew-normally distributed with weakly significant variations over the solar cycle; however, the average magnetic field strength does not display any long term decrease (see Section~\ref{sec:temporal_effects}). 


\section{Observations}\label{sec:obs_reduce}

The SP onboard the \textit{Hinode} spacecraft regularly achieves a polarimetric accuracy of $10^{-3}$ of the incoming intensity when measuring the full Stokes polarized spectra of the Fe I 630.15 and 630.25 nm spectral lines with a spectral sampling of 2.15 pm \cite{kosugi2007,tsuneta2008,ichimoto2008}.  The SP spectrograph slit has a projected length of $164''$ and spatially samples the solar surface every $0.16''$ both along the slit and in the stepping direction, which is restricted to 328'' \cite{tsuneta2008}.  As a proposal driven instrument, the mode of operation and target selection for \textit{Hinode} varies.  All of the spatial SP maps analyzed here have at minimum an effective resolution of $0.64''$, corresponding to a spatial sampling of $0.32''$.  The SP is a stable platform.  Seeing fluctuations are non-existent and platform vibrations do not degrade the quality of the data.  

\subsection{Data Selection and Azimuth Disambiguation}

We analyze 628 \textit{Hinode}/SP area scans, manually selected from the Level 2 data archive between November 2006 and November 2012.  The SP often scans the same region many times during its progression across the solar disk.  Furthermore, larger active-region targets are typically preferred over smaller sunspots within the \textit{Hinode} target selection process, which might introduce a selection bias if all scans are included.  To avoid this, we do not examine every area scan.  Instead, on each UT calendar day, we select an individual scan for each unique targeted region, and attempt to maximize the time interval between scans taken on adjacent UT days.  Scans that have a large field-of-view are preferred in our selection over scans of a reduced field-of-view, which typically were acquired in a higher resolution mode.  We also ensure that the scans have full coverage over each individual sunspot umbra.  Of the selected scans, 72 had a high East-West (x) (North-South (y)) spatial sampling of $0.148''(0.160'')$, 553 had an x(y) sampling of $0.297''(0.320'')$, and 3 had a mixed x(y) sampling of $0.148''(0.320'')$. 

The Level 2 data processing includes inversions of the measured Fe I polarized spectra at each spatial pixel using the {\sc MERLIN} Milne-Eddington inversion code \cite{lites2009}.  {\sc MERLIN} fits the intensity spectrum with a weighted combination of a single magnetic atmospheric component and a stray-light component using the normal definition for the fill fraction.  The magnetic component is characterized by a single magnetic field strength, and can be considered an average value over the range of spectral-line formation in the atmosphere.  The stray-light profile is determined from the average intensity spectrum from regions within the scan with low polarization.  For sunspot umbrae, the magnetic fill fraction is typically high ($> 0.8$).  To study the inclination of each umbral core magnetic field in the local reference frame of the sunspot, we resolve the $180^{\circ}$ azimuth ambiguity of the transverse Zeeman effect within each sunspot umbra for the 159 target scans located near the central meridian ($\pm 10^{\circ}$) using the AZAM disambiguation tool \cite{lites1995} prior to transforming the LOS vector into the local reference frame.  Due to the divergence (convergence) of the sunspot magnetic fields for positive (negative) polarity umbrae, the disambiguation of the umbral fields is a simple, straightforward process. 

\subsection{Automated Threshold-based Sunspot Detection}

\begin{figure}
\centerline{\includegraphics[width=1.\textwidth]{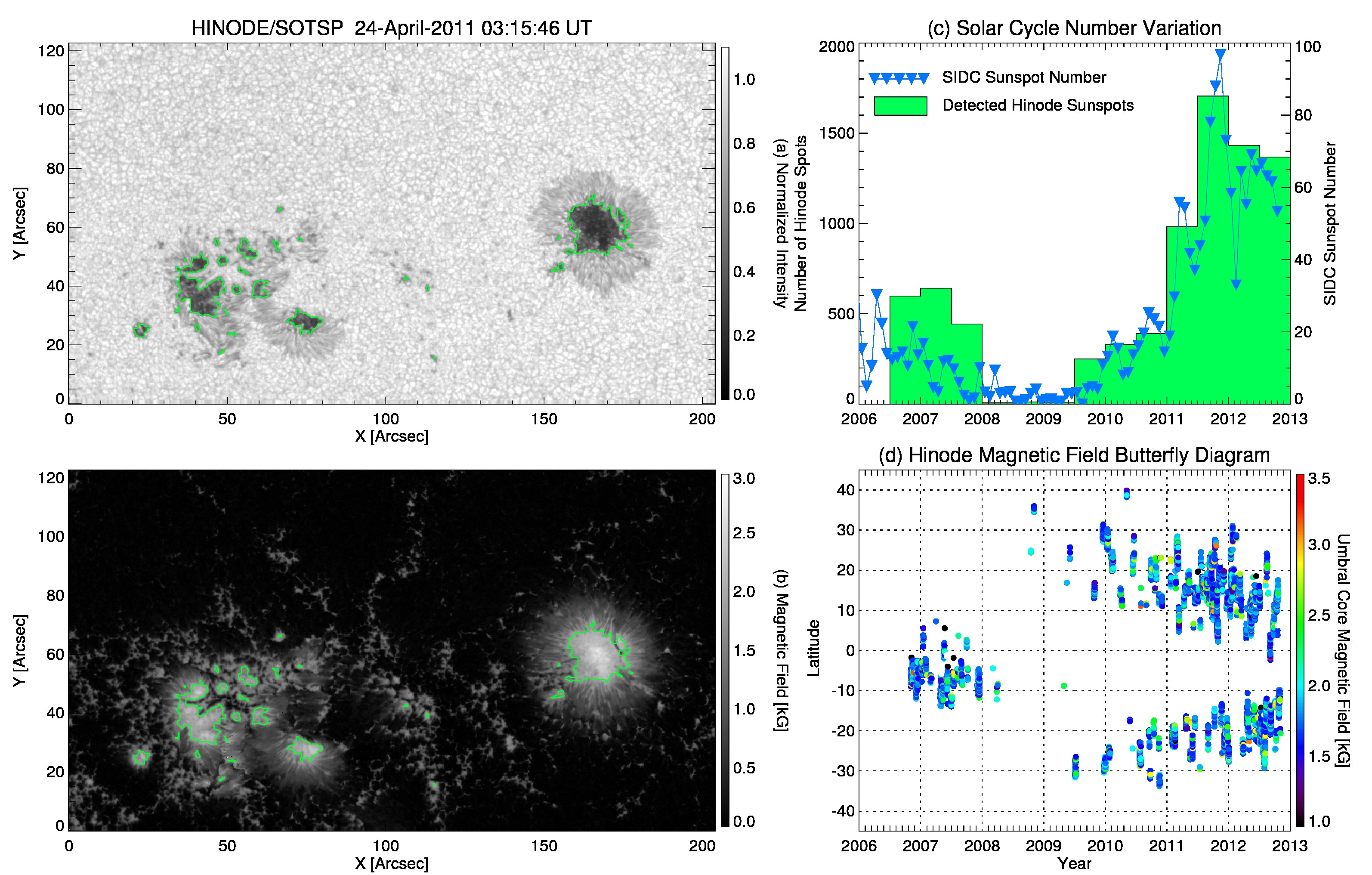}}
\caption{(\textit{top left}) An example of our automated selection of sunspots and pores for one scan from the \textit{Hinode Spectropolarimeter} Level 2 data archive. All structures with a normalized intensity below 0.575 are identified as a sunspot feature.  The total magnetic field (\textit{bottom left}) is recorded in these Level 2 data from a Milne-Eddington inversion, which includes weighting by a non-magnetic filling factor determined directly from the polarized spectral fits. The yearly distribution of all 7530 selected spots adequately samples the latitudinal dependence of sunspot emergence through the solar cycle as shown in the butterfly diagram of the core magnetic field strength (\textit{bottom right}), while the total number distribution tracks the solar cycle variation in the SIDC international sunspot number (\textit{top right}).}\label{fig:selection}
\end{figure}

A total of 7530 dark features are found in our \textit{Hinode} data set using an automated method based on intensity thresholding of the continuum intensity maps for each scan, as in \inlinecite{schad2010}.  These dark features consist of sunspot umbrae as well as pores without penumbra, which we collectively refer to as simply umbrae.  The continuum intensity of each scan is first corrected for the center-to-limb dependence of the continuum intensity at 630 nm using the wavelength-dependent values given by \inlinecite{pierce2000}.  The intensity is then normalized over the entire scan by the 75th percentile value of the intensity; that is, we divide by the value of the center-to-limb corrected intensity that separates the upper 25\% of the intensity values from the lower 75\% values.  We find this a better automated method for normalization due to the varying amount of plage.  Regions identified as umbrae are required to have a normalized intensity value below 0.575.  Dilation and erosion image processing techniques are employed to connect neighboring regions separated by a small gap. Every selected feature must also contain at least 4 observed spatial pixels.  An example of the automated selection is presented in Figure~\ref{fig:selection}.

The slit-scanning mirror responsible for scanning the solar image across the \textit{Hinode}/SP spectrograph slit does not translate linearly \cite{centeno2009}.  Instead, it periodically deviates from a linear translation with an amplitude of about one arcsecond.  While this small error is not expected to introduce large errors in the location of the darkest portion of the umbra from which we glean parameters, it does influence the determined umbral sizes.  We correct the non-linear scanning using the calibrated map labeled by the `X\_coordinate' keyword within the Level 2 data.  Spatial masks recording the identified umbrae are destretched to the median scanning step prior to measuring umbral sizes.  

In total, 1117 umbrae are found in the high resolution maps, 6408 in the lower resolution maps, and 5 in the mixed resolution maps.  We record the location, normalized intensity (or contrast), magnetic field strength, LOS inclination, and LOS azimuth for both the darkest pixel in each umbra as well as the pixel with the greatest magnetic field strength.  We find a good correlation between the location of the largest field strength and darkest intensity.  Thus, we limit our study only to the pixels with the darkest continuum intensity within each umbra in accordance with previous studies \cite{penn2006,penn2007,schad2010}.  The total area of each umbra is determined by the number of spatial pixels it covers and the foreshortened-corrected effective area of each pixel in physical units.  The radius of each umbra is then calculated assuming the umbra is circular.  

As shown in Figure~\ref{fig:selection}c, the total number of spots found in our data selection varies with the solar cycle in accordance with the \textit{Solar Influences Data Analysis Center} (SIDC) sunspot number \cite{sidc}.  The magnetic butterfly diagram in Figure~\ref{fig:selection}d gives the core magnetic field strength and latitudinal location of each umbra as a function of the solar cycle.  Low latitude umbrae from Solar Cycle 23 are measured between November 2006 and May 2008.  Umbrae from Solar Cycle 24, starting in late 2008, exhibit the familiar equator-ward drift in latitudinal locations as a function of solar cycle.  We conclude from this figure that our selection of spots adequately samples umbrae from the full latitudinal extent of sunspot formation during the solar cycle. 

\subsection{Stray-Light Considerations}

We correct the determined umbral core intensities for instrumental stray light by accounting for the fill fraction values reported directly by the {\sc MERLIN} inversions.  According to \inlinecite{lites2009}, during the inversions the Stokes spectra are fitted with a single magnetic component and a field-free stray-light profile according to:
\begin{equation}
I_{syn}(\lambda,f,{\bf{x}}) = f I_{mag}(\lambda,{\bf{x}}) + (1-f) I_{stray}(\lambda) \label{eqn:ff_fit}
\end{equation}
where $I_{syn}$ denotes the synthetic Stokes profiles fitted to the observed profiles.  These calculated profiles result from the weighted contribution of a magnetized atmospheric component, $I_{mag}$, and a scattered light profile, $I_{stray}(\lambda)$
The properties of the magnetized atmosphere are described by the standard set of parameters, ${\bf{x}}$, including the Milne-Eddington 
factors, thermal line width and damping, and a single magnetic field vector.  The weighting factor, f, is known as the magnetic fill fraction or, alternatively, $(1-f)$ denotes the stray-light fraction.  $I_{stray}(\lambda)$ is determined directly from the data by averaging the profiles within spatial points of low total polarization.  Then, the returned magnetic field already has taken into account the role of stray light, where as our determination of the intensity has not.  We correct the umbral contrast measurements, as in \inlinecite{mathew2004}, by defining, in accordance with Equation (\ref{eqn:ff_fit}), the corrected umbral contrast as:
\begin{eqnarray}
R^{c}_{spot} & = &  \frac{R^{c}_{obs} + (f - 1)R^{c}_{stray}}{f} \approx \\ \nonumber 
                       & \approx & \frac{R^{c}_{obs} + (f - 1)}{f}  \label{eqn:int_correct}
\end{eqnarray}
where $R^{c}_{obs}$ is the umbral intensity contrast measured in the continuum, determined directly from the \textit{HINODE}/SP intensity maps that have been flattened and normalized.  We make the approximation that the light scattered into the umbra, $R^{c}_{stray}$, originates from the quiet Sun, such that $R^{c}_{stray} = 1$.


\section{A Continuum of Thermal and Magnetic Field Properties}\label{sec:B_I_TEMP}

\subsection{Relationship of Umbral Contrast to Magnetic Field Strength}

\begin{figure}
\centerline{\includegraphics[width=1.\textwidth]{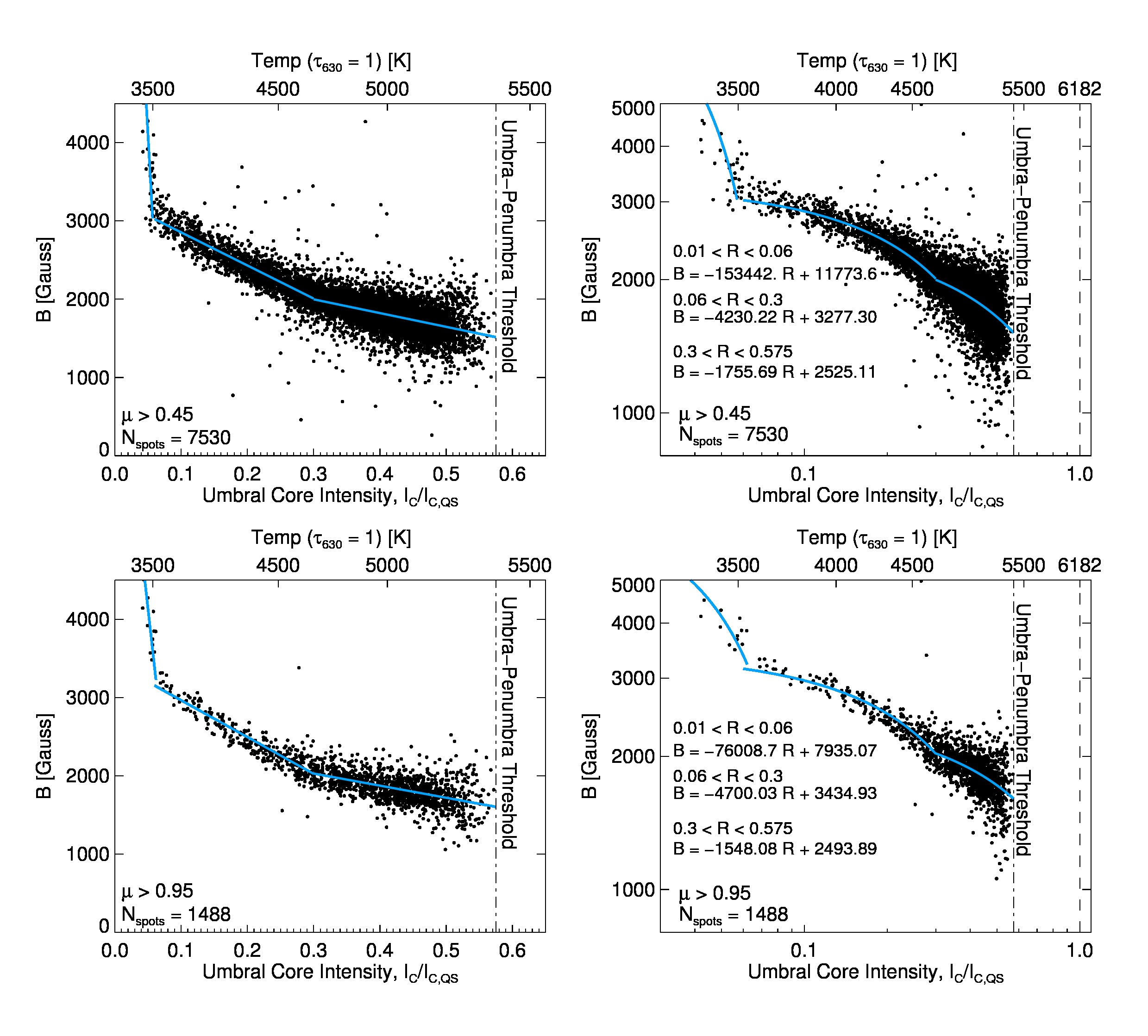}}
\caption{Magnetic field strength \textit{versus} normalized intensity for dark umbral cores.  The top panels show the relationship for all the identified spots, while the bottom panels show the relationship for umbrae measured near disk center.  Left and right plots correspond to linear-linear and log-log representations of the same data.  Linear fits (blue continuous lines) of the observed data are over-plotted, and quantified within the figure for three ranges of the normalized intensity (denoted by R).  Note that the magnetic field strengths for the darkest umbrae (R$ < 0.06$) are anomalously high due to the effect of molecular blends on the Stokes inversion process.  The temperature scale given at the top of each panel refers to the umbral core temperature at unit continuum optical depth inferred using Planck's law. }\label{fig:int_mag}
\end{figure}

\inlinecite{king1934} first discovered the strong correlation between the peak magnetic field strength and the photometric intensity of a collection of sunspots; furthermore, he discussed the wavelength dependence of this correlation.  Since then, the empirical relationship between the emergent intensity and the magnetic field strength has since been studied by a number of authors both  within individual sunspots \cite{martinez_pillet1993,solanki1993,leonard2008,jaeggli2012} and amongst a collection of sunspots \cite{kopp1992,penn2003,norton2004,schad2010,rezaei2012}.  Collections of sunspot umbrae observed by the SOHO/MDI \cite{norton2004} and the KPVT/SPM \cite{schad2010} exhibit power law relationships between the minimum umbral intensity and the corresponding magnetic field strength.  Yet, these studies are hindered by low number statistics (\textit{esp.} \inlinecite{norton2004}), and scatter-inducing instrumental effects (\textit{e.g.} stray-light and magnetograph ``saturation'' effects).  

Taking advantage of the excellent photometric and spectropolarimetric properties of the SP, we re-investigate the relationship between intensity and magnetic field strength for our collection of 7530 identified umbral cores (see Figure~\ref{fig:int_mag}).  The relationship between the magnetic field strength and the continuum contrast is shown both for all of the umbrae (top panels in Figure~\ref{fig:int_mag}, $\mu > 0.45$, where $\mu$ is the cosine of the umbra's heliocentric angle) and for the spots close to disk center (bottom panels in Figure~\ref{fig:int_mag}, $\mu > 0.95$).  Evidenced by the lack of linearity in the right log-log scaled plots, a single power law relationship fails to accurately characterize these data.  Instead, we empirically describe the trends \textit{via} three linear fits for the ranges of R indicated in the figure.   Furthermore, the scatter in these data is greatly reduced in comparison with previous studies.  Since the SP resolves individual umbral dots, we expect our ability to locate the darkest portion of the umbra to be improved, which, in addition to a reduced stray light, may explain a lower scatter.  

For high-contrast spots, a pronounced, nearly-isothermal strengthening of the magnetic field amongst the darkest umbral cores is evident.  However, while such behavior has been observed within individual spots using the infrared Fe I lines at 1565 nm \cite{jaeggli2012}, the behavior seen here is probably a result of the influence of TiO and MgH molecular lines on the {\sc MERLIN} inversions.  We extract the Hinode Stokes spectra from the 29 scans with an umbra exhibiting a normalized intensity below 0.1 and a magnetic field strength above 3250 G.  This subsample contains only 12 unique umbrae.  We perform independent Milne-Eddington inversions on the spectra of the darkest umbra.  The results return inconsistent measures of the magnetic field.  Though, in general, the inferred magnetic field strengths for these umbrae are reduced in relation to the Level 2 \textit{Hinode} values.  It is clear that these spectra must be analyzed with more appropriate atmospheric models incorporating molecular formation and the molecular Zeeman effect (see, \textit{e.g.}, \inlinecite{wenzel2010} and \inlinecite{berdyugina2011}).

\subsection{Temperature - Magnetic Field Relationship}

\begin{figure}
\centerline{\includegraphics[width=1.\textwidth]{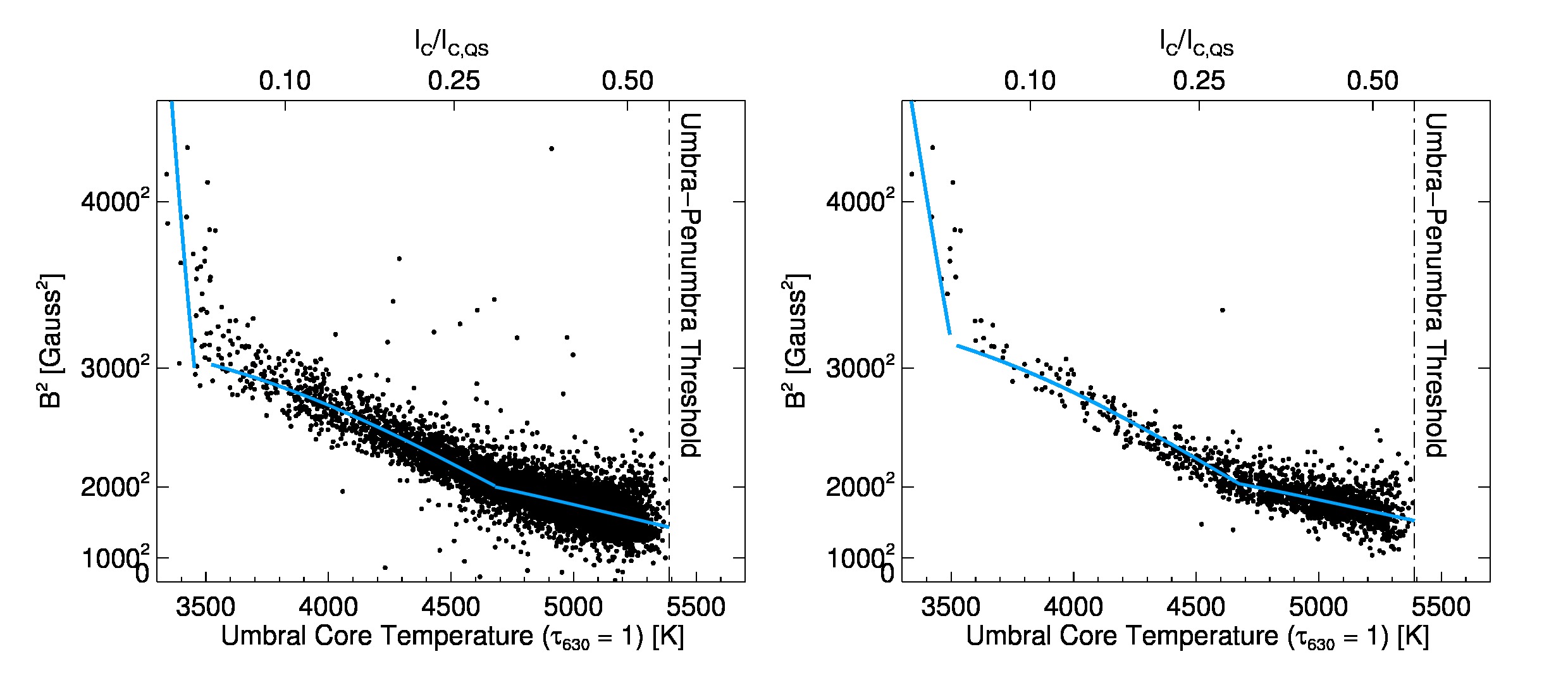}}
\caption{Squared magnetic field strength \textit{versus} umbral core brightness temperature for all the measured umbrae ($\mu > 0.45$, left panel), and for those umbrae close to disk center ($\mu > 0.95$, right panel).  The linear fits found in Figure~\ref{fig:int_mag} are over-plotted, using Equation (\ref{eqn:planck_temp}) to convert normalized intensities into brightness temperatures.}\label{fig:temp_sqmag}
\end{figure}

A theoretical understanding of the intensity-magnetic field strength relationship requires a radiative magneto-hydrodynamic (MHD) approach akin to the studies of \inlinecite{rempel2009} and \inlinecite{borrero2010}.  Yet, a simple picture can be built from the static force balance consisting of the external gas pressure from the photosphere/upper convective zone, the internal gas pressure of the sunspot, and the horizontal component of the magnetic Lorentz force.  Following the assumptions of \inlinecite{maltby1977} and the notation of \inlinecite{solanki1993}, the horizontal magnetohydrostatic (MHS) pressure equation can be written in cgs units as: 
\begin{equation}
P_{0}(z) = P_{g}(r,z) + \frac{B_{z}^{2}(r,z) + F_{c}(r,z)}{8\pi}, \label{eqn:mhs_eq}
\end{equation}
where
\begin{equation}
F_{c} = 2 \int_{r}^{a} B_{z}(r',z) \frac{\partial B_{r}(r',z)}{\partial z} dr'.
\end{equation}
$P_{0}$ refers to the external gas pressure, while $P_{g}$ is the internal gas pressure of the sunspot, $r$ is a radial coordinate measured from the sunspot center in a cylindrical geometry with $z$ being the height coordinate.  The Lorentz force is decomposed into the two terms corresponding to a pressure term and the field curvature tension term, $F_{c}$.  As in \inlinecite{jaeggli2012}, under the limited assumptions of a vertical magnetic field in a field-free, ideal background atmosphere of constant height and density, Equation (\ref{eqn:mhs_eq}) reduces to a simple proportionality between the squared magnetic field strength and the difference between the external and internal temperatures, \textit{i.e.}
\begin{equation}
B^{2} \propto T_{qs} - T_{s} \label{eqn:mhs_simple}
\end{equation}
Accordingly, we display in Figure~\ref{fig:temp_sqmag} the relationship between the squared magnetic field strength and the umbral core temperature determined from our collection of observed umbral cores.

The umbral core temperature is determined from the normalized continuum intensity and the Planck function, as in \inlinecite{solanki1993}, using the equation
\begin{equation}
R^{c}_{spot} = \frac{I^{c}_{spot}}{I^{c}_{QS}} = \frac{e^{hc/\lambda k_{B}T_{0}} - 1}{e^{hc/\lambda k_{B}T} - 1},\label{eqn:planck_temp}
\end{equation}
where $h$, $c$, $\lambda$, $k_{B}$ take their customary meanings.  $I^{c}_{spot}$ and $I_{QS}^{c}$ are the intensities determined in the spectral continuum within the sunspot and the local quiet Sun, respectively.  $T$ and $T_{0}$ are the brightness temperatures of the spot and of the quiet photosphere.  $T_{0}$ is assigned for our 630 nm channel to $6182$ K, in accordance with the measurements in \inlinecite{maltby1986}. 

Just as in the studies of the temperature-magnetic field strength relationship within individual sunspots, dark umbral cores when viewed in collection do not comply with the simple picture of MHS balance given by Equation (\ref{eqn:mhs_simple}).  As shown in Figure~\ref{fig:temp_sqmag}, while the temperature and squared magnetic field strength are clearly well correlated, the relation is not linear.  Once again, the strengthening of the umbral field for umbrae with low temperatures produces a pronounced effect; although, this is probably an artifact induced by the molecular formation (see the discussion in the previous section).  Furthermore, umbrae with higher temperatures exhibit a non-linear relation.  As particularly apparent in the disk center sampling of umbrae, the slope of the $B^{2}$ \textit{vs.} $T$ relationship varies between intermediate and warmer temperatures, which is not found in other studies.  It is unclear at this time what generates this difference; though, we note that we do not distinguish in our sample between umbrae with and without penumbra.  The smaller, brighter features that we measure probably have no penumbra, while the larger, darker features do.  The presence of a penumbra can significantly alter the force balance throughout a magnetic field concentration, which might influence the $B^{2}$ \textit{vs.} $T$ relationship.  Furthermore, the formation height of the continuum within sunspot umbrae is influenced \textit{via} the well-known Wilson depression effect by the magnetic field intensity.  Consequently, the simple proportionality of Equation (\ref{eqn:mhs_simple}) no longer holds, leading to changes in the observed $B^{2}$ \textit{vs.} $T$ relationship.  Measurements of the Wilson depression depth can be used to constrain the curvature force within an umbraa, but such measurements are difficult to achieve.


\section{Umbral Size Dependencies}\label{sec:size_depend}

\begin{figure}
\centerline{\includegraphics[width=1.\textwidth]{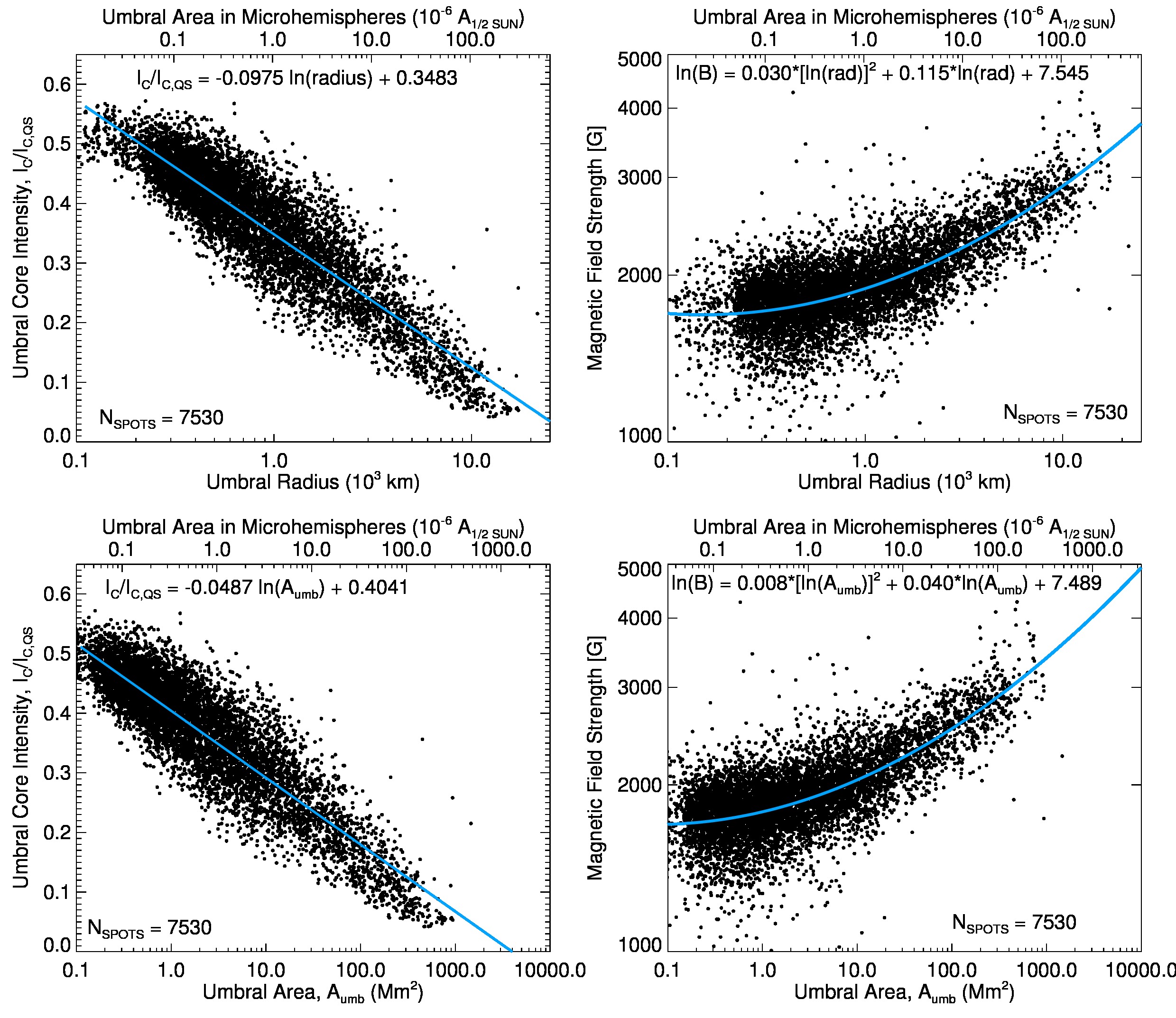}}
\caption{Dependence of the umbral core intensities (left panels) and the total magnetic field strengths (right panels) on the umbral radius (top panels) and umbral area in square megameters (bottom panels).  Linear-log and log-log regression curves (blue continuous lines) provide empirical fits to the data points.  The equations within the panels are in the same units as the left and bottom axes of each plot.  As a reference, umbral areas are given in units of microhemispheres in the top axes of each panel.}\label{fig:mag_int_size}
\end{figure}

In Figure~\ref{fig:mag_int_size}, the magnetic field strength and normalized continuum intensity measured in the umbral cores are related to the overall umbral size, quantified both as a radius (\textit{i.e.} an ``effective radius" assuming a circular umbra) and the total fore-shortened corrected area.  Even though the \textit{Hinode}/SP maps are of considerably higher resolution than the data used in previous determinations of the temperature-magnetic size dependence, considerable scatter still exists can be seen in the figure.  The complex geometry of many umbral features is no doubt at the origin of this scatter.  A regressive fit is shown for each relationship in Figure (\ref{fig:mag_int_size}).  We emphasize that umbral size (as well as intensity and magnetic field strength) depends on the definition of the umbra-penumbra boundary intensity and the wavelength of observation.  Yet, it is clear from these observations that the relationship between the umbral size and the thermal-magnetic properties of the umbra is not linear, which has interesting consequences below (see Section \ref{sec:temporal_effects}).

\section{Deviation from the Vertical Field Assumption}\label{sec:vector_distrib}

\subsection{How vertical are magnetic fields in umbral cores?}

The angle between the magnetic field vector and the LOS direction is quantified as the magnetic field inclination angle (in the LOS reference frame).  Zeeman diagnostics deterministically return the inclination angle \textit{via} the analysis of the Stokes profiles.  Unlike the azimuthal angle (\textit{i.e.} the projection of the magnetic field vector on the plane perpendicular to the LOS direction), the inclination angle determination is unambiguous.  Yet, this inclination angle is measured with respect to the LOS, not to the local solar reference frame.  In order to determine the inclination in the solar reference frame, a transformation, which involves not only the inclination angle but also the azimuthal one, must be performed.  Thus, the determination of the local inclination angle of the magnetic field is subject to our ability to resolve the azimuthal angle ambiguity in the LOS frame.  

\begin{figure}
\centerline{\includegraphics[width=1.\textwidth]{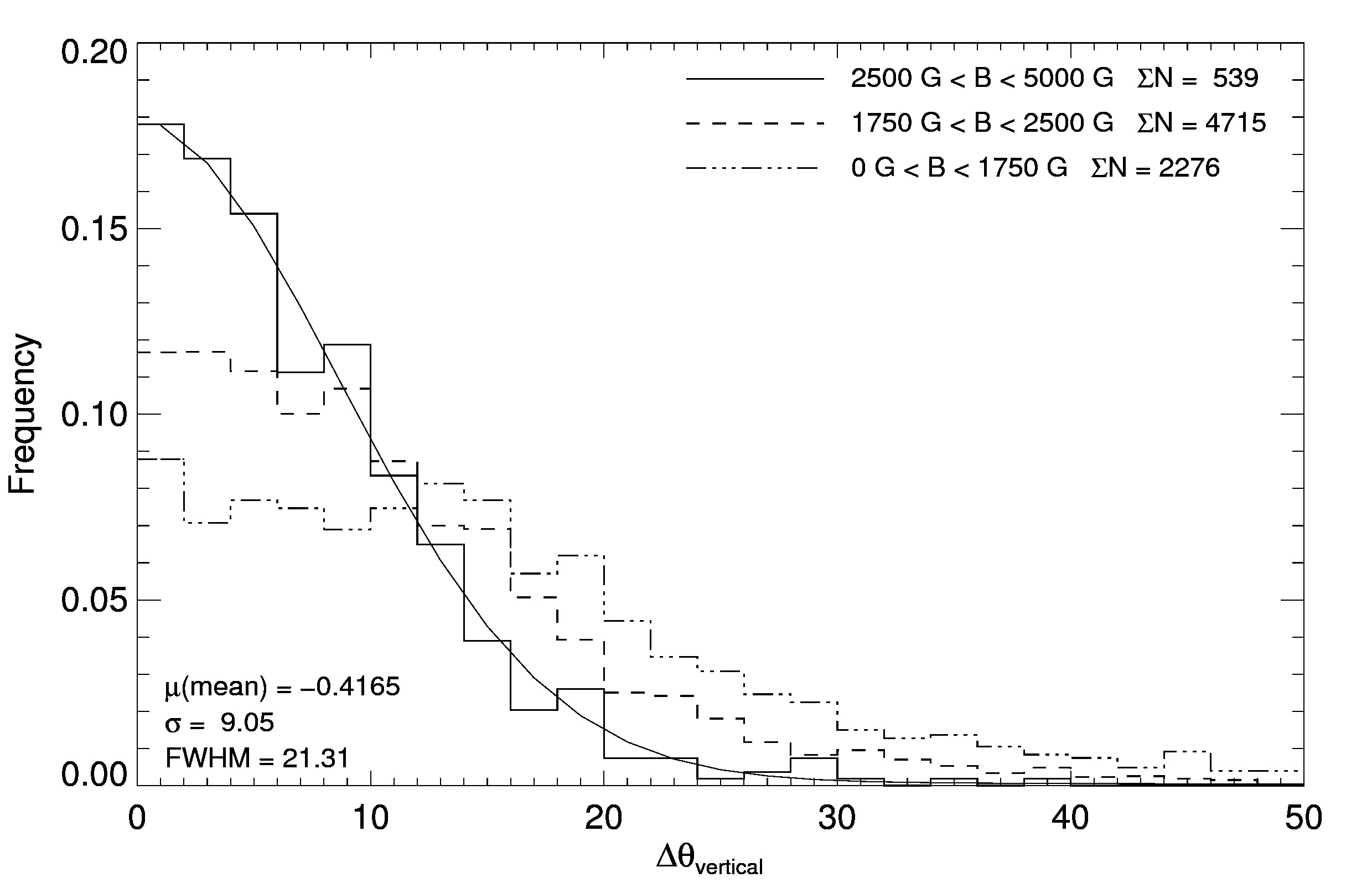}}
\caption{Histograms of the unsigned deviation of the umbral core magnetic field inclination from the local vertical axis.  As expected, weaker umbral cores present a larger deviation range. $\Sigma N$ give the total number of umbrae included within each histogram.}\label{fig:vert_dev}
\end{figure}

For a quick estimation of the deviation of the umbral magnetic field vector orientation from the local solar vertical, one can calculate the deviation of the measured inclination angle in the LOS coordinate system from the expected inclination angle of the magnetic field in this same reference system, assuming a perfectly vertically directed field vector.  This method evaluates the deviation of the core umbral magnetic fields from the vertical direction without the need for azimuth disambiguation.  This quantity can be written as
\begin{equation}
{\Delta{\theta_{vertical}} = \left | \arccos{\left(\mu\right)} - \arccos{\left( \frac{\left|B\cos{\theta_{B}}\right|}{B} \right )} \right |},\label{eq:delta_theta_vert}
\end{equation}
where $\mu$ is the cosine of the angle between the LOS and the local solar vertical, $B$ is the magnitude of the magnetic field, $\theta_{B}$ is the inclination of the magnetic field in the LOS reference frame that ranges from $0^{\circ}$ to $180^{\circ}$ in these Level 2 \textit{Hinode} inversions.  The second term in Equation (\ref{eq:delta_theta_vert}) may be thought of as an ``unsigned LOS magnetic field inclination angle'', whose range is between $0^{\circ}$ and $90^{\circ}$.  Histograms of this deviation angle are given in Figure~\ref{fig:vert_dev}.  The one-sigma deviation from the vertical field approximation for strong umbral cores is $\sim 9^{\circ}$, confirming earlier results of vertically directed fields in strong umbra.  The vertical field approximation quickly vanishes for weaker umbrae. 

\subsection{Joy's Law and the Vector Magnetic Field of Umbral Cores}\label{sec:vec_distrib}

\begin{figure}
\centerline{\includegraphics[width=1.\textwidth]{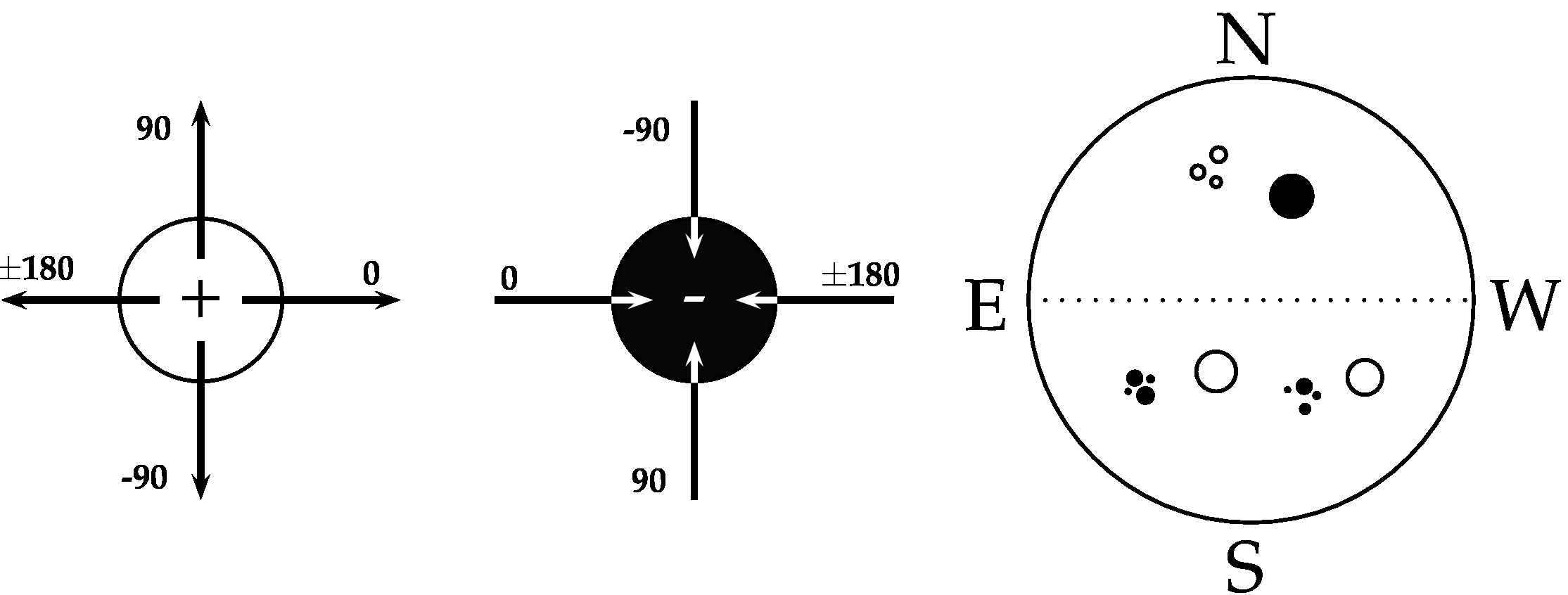}}
\caption{An illustration of the azimuthal angle reference frame used in this work, and Hale's and Joy's law for Solar Cycle 24 (December 2008 - present).  The diagrams on the left show a diverging and converging transverse magnetic field extending from a positive and negative polarity region, respectively.  In the local reference frame of each sunspot, an azimuthal angle of $0^{\circ}$ is directed towards solar West.  Hale's Law, as depicted in the right diagram, predicts that negative polarity spots would lead in bipolar active regions in the northern hemisphere for Solar Cycle 24. }\label{fig:joys_azi_illust}
\end{figure}

Umbral core magnetic fields exhibit a range of deviation angles from the local vertical direction.  We investigate if this deviation might be organized in any way.  \inlinecite{karachik2010} found that the peak magnitude of the average LOS magnetic flux for each polarity of a bipolar active region was measured on separate sides of the solar meridian, putting in evidence that opposite polarity magnetic fields within an active region may be tilted towards each other at the solar surface.  We investigate this possibility using the information of the disambiguated magnetic field vectors of our identified umbrae.  The individual active regions are not investigated separately; rather, we test the key phenomenological laws describing the global organization of the solar magnetic field, \textit{i.e.} Hale's and Joy's laws.  Restricting the investigation only to spots observed after December 2008 within Solar Cycle 24, our sample is separated into four categories based on Hale's polarity law for Solar Cycle 24, as shown in Figure~\ref{fig:joys_azi_illust}, where spots are classified by polarity and hemisphere.  Umbral cores with local inclination angles less than (greater than) $90^{\circ}$ are considered positive (negative) polarity spots.

\begin{figure}
\centerline{\includegraphics[width=1.\textwidth]{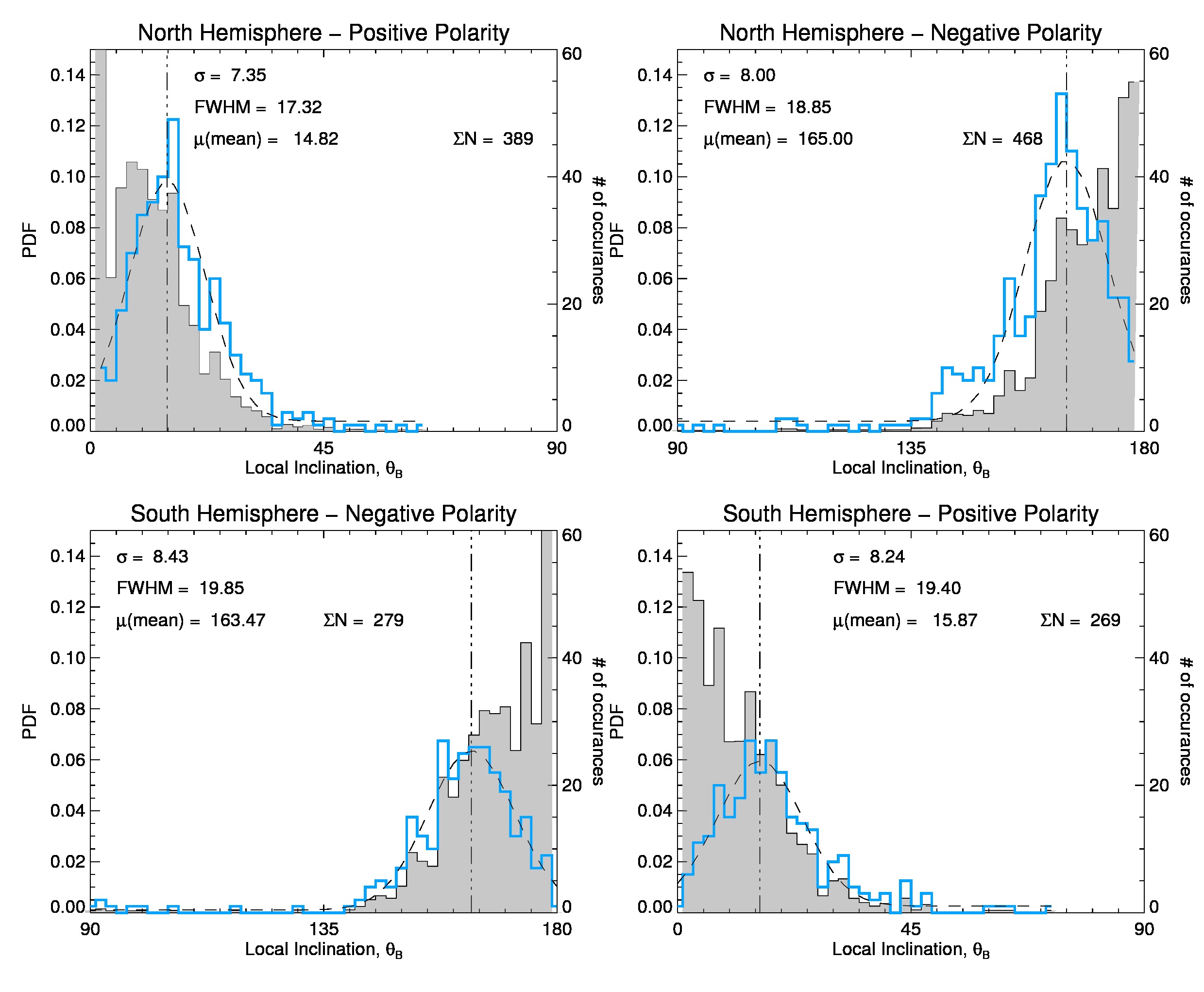}}
\caption{Histograms of the local solar inclination angle of the umbral core magnetic field for umbrae classified according to Hale's polarity law in Solar Cycle 24.  The blue curve gives the raw histogram for $2^{\circ}$ inclination bins, while the grey filled curves reports the probability density weighted by the relative area of each annulus of inclination.  Gaussian functional fits to the raw histograms are performed, with the parameters given in each plot.}\label{fig:hale_inc}
\end{figure}

Histograms of the local solar inclination angle of the umbral cores in the four categories are illustrated in Figure~\ref{fig:hale_inc}.  They are ordered according to the Cycle 24 Hale's polarity law with histograms of the northern hemisphere umbrae at the top.  Umbrae of all sizes and strengths are included.  Two distributions are included in the figure.  The best representation of the number of umbrae with a given local inclination value is given by the regular histogram (blue curve) for $2^{\circ}$ bins.  The parameters describing the fit to these curves by Gaussian distribution functions are given in the figure.  Note that these distributions have peaks away from $0^{\circ}$ and $180^{\circ}$, which correspons to radially directed fields.  This offset \textbf{does not} imply that umbral core magnetic fields are preferentially non-radial. Rather, to calculate a probability density function for the umbral core inclination angles, one must take into account the relative area of each inclination annulus in a spherical geometry.  To do this, we calculate the density function: 
\begin{equation}
PDF(\Delta\theta) = \frac{1}{p} \frac{\Delta N_{\theta}}{\Delta A / A }, 
\end{equation}
where $\Delta N_{\theta}$ corresponds the number of umbrae within a bin of inclination angles between $\theta_{0}$ and $\theta_{1}$.  The quantity $\Delta A / A $ gives the relative area of each inclination annulus given by the integral calculus for a unit sphere as
\begin{equation}
\Delta A / A = 2\pi ( - \cos \theta_{1} + \cos \theta_{0} ) / 4\pi = ( - \cos \theta_{1} + \cos \theta_{0} )/2
\end{equation}
Finally, to normalize the curve by the area under the curve, we divide by $p$, which is the total sum of  $\Delta N_{\theta}/ \Delta A / A $ for all $\theta$s.  These distributions, given in Figure~\ref{fig:hale_inc}, confirm the preference for umbral core magnetic fields to be radial; though, they also show the large spread of inclination angles about the radial direction. 

\begin{figure}
\centerline{\includegraphics[width=1.\textwidth]{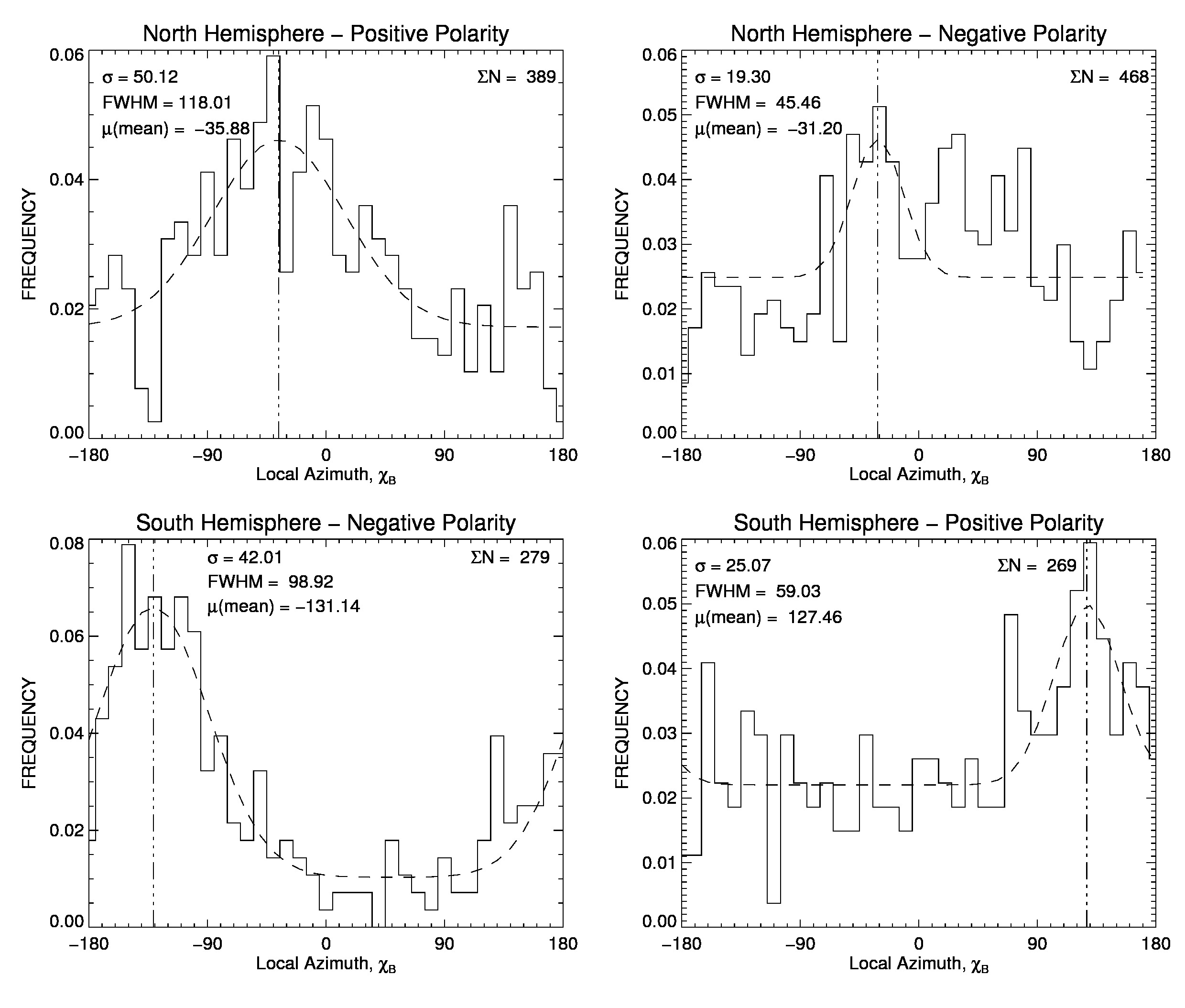}}
\caption{Histograms of the local azimuthal angle of the umbral core magnetic fields for the same selection as in Figure~\ref{fig:hale_inc}.  The azimuthal angles are defined as in Figure~\ref{fig:joys_azi_illust}.}\label{fig:hale_azi}
\end{figure}

Figure~\ref{fig:hale_azi} shows the histograms of the disambiguated magnetic azimuthal angles for the four categories.  An azimuthal angle of $0^{\circ}$ describes a vector directed towards solar west, as indicated in Figure~\ref{fig:joys_azi_illust} for positive and negative polarity spots with a diverging and converging magnetic field, respectively.  While the later histograms exhibit a large spread, the umbral core magnetic fields of the trailing spots exhibit a slight preference to be located following Joy's law.  In the northern hemisphere, the trailing, positive-polarity spots have a mean azimuthal angle of $-35.88^{\circ}$, consistent with a divergent field directed towards the South-West.  Likewise, the trailing, negative-polarity spots in southern hemisphere are consistent with a converging field directed away from the North-West direction.  The leading polarities are not clearly located following Joy's law.  We speculate that these relationships are influenced by the asymmetric distribution of umbrae in the leading versus trailing polarities; that is, the leading polarity flux of a bipolar active region is typically more concentrated than the trailing polarity flux \cite{fan1993}.


\section{Distributed Umbral Properties During The Solar Cycle}\label{sec:temporal_effects}

The accurate characterization of the mean properties of sunspots requires a rigorous analysis of the distribution of these properties.  Historically, this has not been a simple task due to a number of reasons: 1) the lack of instrumental stability and/or the variation of observing conditions, 2) selection bias introduced by manual sunspot identification and/or poor sampling, 3) low number statistics at solar minimum, and 4) the evolution of solar features.  \inlinecite{bogdan1988} overcame the low number statistics by averaging over all solar cycles between 1917 and 1982 measured at Mount Wilson, and identified a log-normal distribution of umbral areas, whose average shape does not change over the solar cycle. Data from the Royal Freenwich Observatory support this log-normal distribution \cite{baumann2005}.  Despite not being a full disk instrument, the \textit{Hinode}/SP data benefits from its stable space-based platform, and our automated analysis eliminates user bias in selecting the umbrae.  As long as our umbral size distribution compares well with historical measurements, we argue our distribution of core magnetic field strengths represents the most accurate determination to date.  Previous distributions of umbral core magnetic field strengths have been presented by \inlinecite{livingston2006}, \inlinecite{rezaei2012}, and \inlinecite{livingston2012}.

We characterize the distribution of umbral property `x' by its probability density function (PDF), $\phi(x)$, which for a continuous population is defined as the derivative of the cumulative distribution function (CDF), $D(x)$ .  Subject to the normalization requirement, the CDF states that 
\begin{equation}
\lim_{x \rightarrow \infty} D(x) = \lim_{x \rightarrow \infty} \int_{-\infty}^{x} \phi(x) dx \equiv 1.
\end{equation}
Understanding the CDF as the probability that `x' is less than some real-value within a population of size N, we can define the discretized CDF, $D'(x)$, as 
\begin{equation}
\lim_{x \rightarrow \infty}  D'(x) = \lim_{x \rightarrow \infty}  \sum_{\mathrm{bins } < x} \frac{1}{N} \frac{\Delta N}{\Delta x} \Delta x \equiv 1,
\end{equation}
such that the discretized PDF takes the form:
\begin{equation}
\phi'(x) =  \frac{1}{N}\frac{\Delta N}{\Delta x}. 
\end{equation}
$\Delta N$ is the number of samples within a bin of width $\Delta x$ centered on x.  We determine $\phi'(x)$ by counting the number of samples within a determined bin size and dividing that number by the bin size and the total number of samples in the population.  This is similar to the umbral area spectrum definition given by \inlinecite{bogdan1988}, except that we have introduced the normalization requirement.  To compare our umbral area PDF with Bogdan's \textit{et al.}, we divided their distribution functions by their scaling factor, $f$, and the number of samples in their umbra population.  

\begin{figure}
\centerline{\includegraphics[width=1.\textwidth]{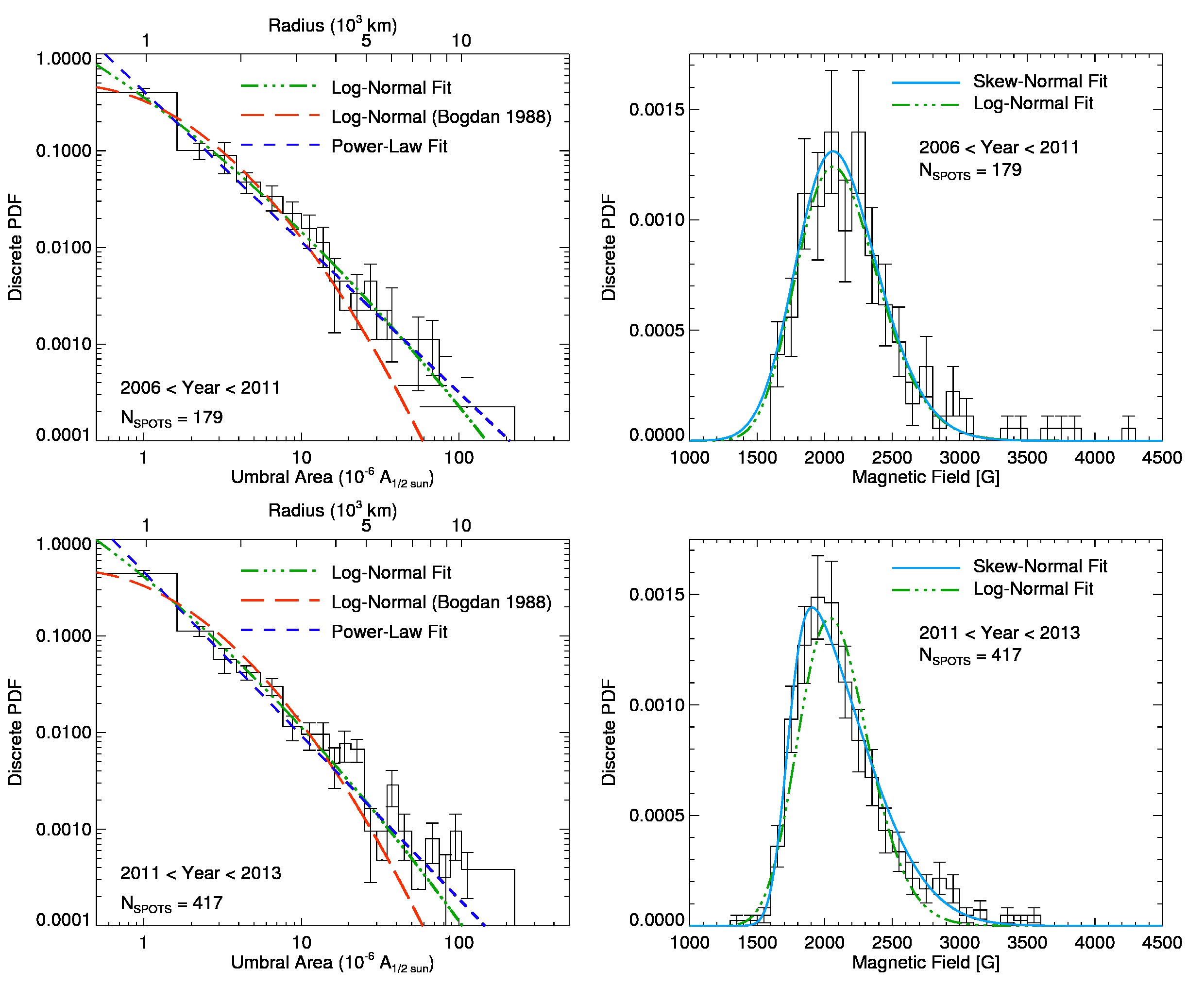}}
\caption{Discrete probability density functions (PDF) of umbral sizes (left panels) and core magnetic field strengths (right panels).  The top plots correspond to the time period between November 2006 and December 2010, while the time period of the bottom plots is January 2011 to November 2012.  Continuous PDF fits are provided as indicated.  See Table~\ref{tbl:mag_size_dist} for fit parameters for these time periods and for all the data.}\label{fig:size_mag_dist}
\end{figure} 

The discretized PDFs for umbral areas (given in units of microhemispheres) and core magnetic field strengths (in cgs units) in our \textit{Hinode}/SP sample are presented in Figure~\ref{fig:size_mag_dist}.  Error bars representing the standard error ($\sigma_{x} = \sqrt{\Delta N}/N/\Delta X$) are shown.  The data are further split into two different time periods, November 2006 to December 2010 and January 2011 to November 2012, which gives some indication of the distributions as a function of solar cycle amplitude.  Using the regression analysis tool {\sc MPFIT} \cite{markwardt2009}, functional forms for the continuous PDFs are fit to the observed discrete PDFs and 1$\sigma$ parameter errors are derived.  The goodness of fit is described by the normal reduced chi-squared parameter, $\chi^{2}$.  To compare with \inlinecite{bogdan1988}, a log-normal function is ascribed to the umbral area distribution, using a similar form, \textit{i.e.}
\begin{equation}
\ln\left(\frac{1}{N} \frac{\Delta N}{\Delta x} \right ) = - \frac{(\ln a - \ln \langle a \rangle)^2}{2 \ln \sigma_{x}} + \ln \left( \frac{1}{N} \frac{\Delta N}{\Delta x}\right )_{max}.
\end{equation}
Additionally, we fit a power law ($\phi'(x) = Cx^{k}$) to the observed umbral area PDFs.  All function parameters and errors are given in Table~\ref{tbl:mag_size_dist}. We find that the distribution of umbral sizes compares well with \inlinecite{bogdan1988} for both time periods, in addition to the full population.  The log-normal description works well, with slightly better fits than using power laws for all time periods.  We do not expect the distribution to match perfectly with that of \citeauthor{bogdan1988}, since \textit{Hinode} probes a different portion of the solar continuum and our umbra-penumbral intensity definition is likely different.  Yet, the log-normal shape of our distribution confirms that our selection is an adequate representation of the distribution of umbral parameters on the Sun.  

\begin{table}
\caption{Parameters for Umbral Magnetic Field Strength and Size Distributsion.  Parameter errors are shown between brackets in the second, third, and fourth columns.}\label{tbl:mag_size_dist}
\begin{tabular}{lcccc}
\hline
Core Magnetic Field & & & & \\ 
 \hline
 \bf{Log-Normal}    & & & & \\      
 Time Period ($N_{spots}$)  &   $<$A$>$      &   $\sigma_{A}$      &   $(1/N)(\Delta N/\Delta a)_{max}$  &   $\chi^{2}$    \\  
2006 - 2011 (179)    &  2062.64 (28.39)  &   1.0211 (0.0033)  &   0.0012 (0.0001)  &   0.78 \\
2011 - 2013 (417)  &  2040.97 (13.42)  &   1.0161 (0.0012)  &   0.0014 (0.0001)  &   1.78  \\
2006 - 2013 (596)      &  2065.67 (11.49)  &   1.0163 (0.0010)  &   0.0014 (0.0001)  &   2.21  \\
 \hline
\bf{Skew-Normal}  & & & & \\
 Time Period ($N_{spots}$) & Location $(\xi)$   & Scale $(\omega)$   & Skew $(\alpha)$  &   $\chi^{2}$   \\ 
2006 - 2011  (179)  &  1830.97 (77.49)  &   426.27 (64.1387)  &   1.6229 (1.0056)  &   0.82  \\
2011 - 2013 (417)  &  1722.50 (18.24)  &   501.47 (26.0172)  &   5.0831 (0.9923)  &   0.71  \\
2006 - 2013   (596)     &  1730.31 (15.53)  &   518.59 (22.1018)  &   5.1874 (0.8081)  &   0.87  \\
 \hline
Umbral Areas & & & \\ 
 \hline
 \bf{Log-Normal}     & & & & \\
Time Period ($N_{spots}$)         &   $<$A$>$      &   $\sigma_{A}$      &     $(1/N)(\Delta N/\Delta a)_{max}$     &   $\chi^{2}$    \\  
2006 - 2011 (179)         &   0.0014 (0.0056)   &   262.37 (716.11)   &  17.813 (54.250) &   0.49  \\
2011 - 2013 (417)                   &   0.0011 (0.0032)   &   170.74 (319.25)   &  39.404 (97.371) &   1.45  \\
2006 - 2013   (596)            &   0.0012 (0.0027)   &   214.20 (320.80)   &  27.300 (50.002) &   1.33  \\
 \hline
\bf{Power Law}  & & & & \\
Time Period ($N_{spots}$)  &   Constant (C)   &   Exponent (k)  &  -   &   $\chi^{2}$  \\ 
2006 - 2011 (179)  &    0.4126 (0.0396)   & -1.5582 (0.0522)  & - & 0.58  \\
2011 - 2013 (417)  &    0.4587 (0.0283)   & -1.6920 (0.0386)  & - & 1.49  \\
2006 - 2013 (596)  &    0.4442 (0.0229)   & -1.6320 (0.0299)  & - & 1.46  \\
\hline 
\end{tabular}
\end{table}

The PDF of the core magnetic field strength should be the transformation of the size PDF subject to the relationship between size and magnetic field discussed in Section~\ref{sec:size_depend}.  Previous investigations of the umbral area-magnetic field relationship have found a linear relationship \cite{rezaei2012}; though, our high-resolution data displays a relationship that deviates somewhat from linearity.  If the relationship is linear, one would expect a log-normal distribution in the core magnetic field strengths.  Since our relationship is not linear, we seek an additional functional form to describe the magnetic field PDF.  We introduce the skew-normal function described by \inlinecite{azzalini1985class}:
\begin{equation}
f(x,\alpha) = 2\phi(x)\Phi (\alpha x),
\end{equation}
where $\phi(x)$ and $\Phi(x)$ are the standard normal (Gaussian) probability density and cumulative distribution functions defined as
\begin{eqnarray}
\phi(x) & = & \frac{1}{\omega\sqrt{2\pi}} e^{-\frac{(x-\xi)^{2}}{2\omega^{2}}} \\
\Phi(\alpha x) & = & \int_{-\infty}^{\alpha x}\phi(t)dt = \frac{1}{2}\left [ 1 + \mathrm{erf} \left (\frac{\alpha(x-\xi)}{\omega\sqrt{2}} \right )\right ]
\end{eqnarray}
Through this definition, the normal parameters describing the Gaussian density function have different meanings.  $\xi$, $\omega$, $\alpha$ are known as the location, scale, and shape (or skew) parameters, respectively.  When $\alpha = 0$, $f(x,\alpha=0)$ is equivalent to a standard normal Gaussian distribution, where the $\xi$ and $\omega$ give, respectively, the mean and the standard deviation.  Although log-normal fits still fit well to the data ($\chi^{2}<2.21$), skew-normal fits seem to better describe ($\chi^{2}<0.87$) the core magnetic field PDF for all time periods (see Table~\ref{tbl:mag_size_dist}).  Furthermore, the skew parameter gives us an idea of a non-Gaussian behavior.  The skewness of the distribution for the later years ($\alpha = 5.08$) is higher than that of the earlier years ($\alpha = 1.62$) in our sample; perhaps due in part to the combined effect of the nonlinear relationship between magnetic field strength and size and the greater raw number of larger spots in the years 2011 and 2012.  Each fit supports a non-zero level of skewness in the PDF, as opposed to the results of \inlinecite{rezaei2012} and \inlinecite{livingston2012}.  

Skewness changes in the umbral core magnetic field PDF may imply variations in the mean magnetic field strength over time.  While our time series is short, the skewness parameters indicate a higher skew when the level of activity is greater (\textit{i.e.} when the number of sunspots is larger).  This is consistent with the presumed steady size distribution of umbrae and the nonlinear relationship between size and magnetic field strength.  When the cycle amplitude is larger, an identically shaped size distribution produces a greater number of larger spots, although the \textit{relative} number of these spots does not change significantly.  The increased number of larger spots reduces the level of sampling noise in the upper tail of the magnetic field strength distribution such that the non-linear relationship between umbral size and core magnetic field strength better influences the determined magnetic field strength distribution.  The associated change in the arithmetic mean can be calculated by the expectation value for a skew-normal distribution given by 
\begin{equation}
E[x] = \xi + \omega \left( \frac{\alpha}{\sqrt{1 + \alpha^2}}  \right) \sqrt{\frac{2}{\pi}}.
\end{equation}
The arithmetic mean for the three skew-normal functional fits to the magnetic field PDF give 2120.53, 2115.09, and 2136.60 G, respectively.  Thus, while the magnetic field PDF shape changes, the mean magnetic field does not significantly vary between the earlier and the later years of our sample. 

\begin{figure}
\centerline{\includegraphics[width=1.\textwidth]{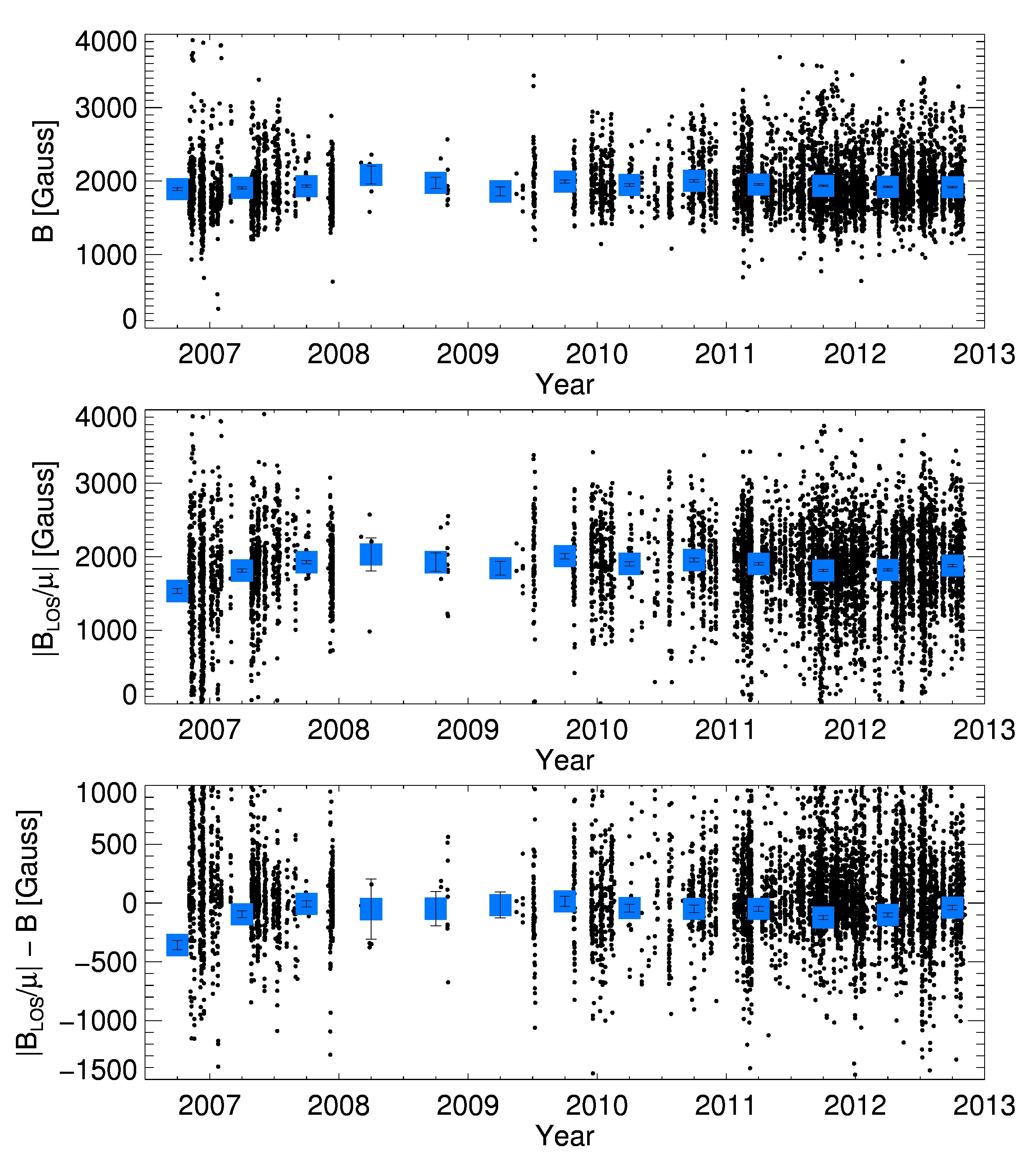}}
\caption{(top) Umbral core magnetic field strength of all 7530 identified umbrae (data points) as a function of time. (middle) The LOS component of the magnetic field divided by $\mu$ (i.e. the vertical (or radial) field approximation) for all spots.  (bottom)  The difference between LOS estimates for the total core magnetic field strength and the true core magnetic field strength.  Blue squares denote the mean field strength for half-year bins, with error bars indicating the $1\sigma$ standard error of the mean.}\label{fig:year_b_blos}
\end{figure}

Finally, we investigate the possibility of long-term secular trends in the mean umbral magnetic field strength, as reported by \inlinecite{penn2011}.  We are also interested in the utility of LOS magnetic flux measurements to investigate trends in the mean magnetic field strength over the solar cycle, when one applies the vertical (or radial) field assumption.  Figure~\ref{fig:year_b_blos} shows the core magnetic field strength and the core LOS approximation for the magnetic field strength, as a function of time, for all the identified umbrae.  The difference of the two measurements is also plotted.  A linear fit to the mean magnetic field strength binned in half-year bins as a function of the year after 2006 (\textit{i.e.}, $(f(year - 2006) = m \times (year-2006) + b$) results in a negligible gradient.  The fitted gradient returns $m = (1.023 \pm 2.13)$ G yr$^{-1}$, with $b = (1929.45 \pm 11.39) G$.   Using the LOS values, the fitted line is described by $m = (5.76 \pm 3.33)$ G yr$^{-1}$, $b = (1827.84 \pm 17.54)$ G.  Compare this with the average gradient shown by \inlinecite{livingston2012} of $(-46 \pm 6)$ G yr$^{-1}$ between  1998 and 2010.  None of our measurements give a long- term decreasing trend.  We do find short-term trends in the data implying possible solar cycle variations, which we claim can occur alongside a steady size distribution.  For example,between July 2009 and November 2012, the mean core magnetic field strength shows a gradient of   $(-24.99 \pm 5.39($ G yr$^{-1}$ while the mean $|B_{LOS}|/\mu$ values display a gradient of $(-39.55 \pm 8.66)$ G yr$^{-1}$.  The LOS measurements clearly overestimate the trend, which is consistent with our discussion on the deviation of the magnetic field vector in dark umbral cores from the strictly radial approximation.  We do not think that the trends shown between July 2009 and November 2012 mean or confirm a long-term decrease in the mean umbral magnetic field strength, \textit{i.e.}, on a time scale longer than a solar cycle.  Rather, it suggests the mean magnetic field strength may vary over the solar cycle.  We find no evidence for a long-term secular decrease.


\section{Discussion}

Our collection of umbrae observed with by the \textit{Hinode}/SP offers the highest resolution account of the thermal and magnetic properties of dark umbral cores to date.  As studies continue to use the collective properties of umbrae to study the variations of the mean sunspot structure and formation over the solar cycle, it is critical to address the complexities involved in relating the various properties of sunspots to the variations of these properties over the solar cycle.  Furthermore, understanding the limitations of the various data sources used is very important. 

First, and foremost, we claim that our selection of umbrae adequately characterizes the full spectrum of umbrae on the Sun, despite the fact that the \textit{Hinode}/SP is not a full-disk instrument.  We claim this because the determined PDF of umbral areas is consistent with previous studies, both for the whole cycle and for different portions of the solar cycle.  Although \inlinecite{nagovitsyn2012} suggest that the distribution of the total sunspot areas (including penumbrae) varies with time, we find no support for changes in the umbral area PDF with time.  The classification and identification of penumbra in a synoptic data set is quite complicated, as penumbrae associated with different umbrae are often merged and/or too close to each other.  The identification of individual umbrae is more straight-forward and reliable.

Both the normalized intensity and the magnetic field strength measured in the darkest region in each umbra show a clear correlation with the total size of the umbra.  The relationships are not linear.  Consequently, the distribution of these parameters are not expected to follow the same distribution as the umbral areas.  Furthermore, while the umbral area PDF remains constant, its amplitude varies as a function of solar cycle phase and as a function of the total solar cycle amplitude.  This leads to a lower or greater number of large spots from cycle to cycle (with no change in the \textit{relative} number of these spots).  This, we expect, explains the variations in the maximum magnetic field strength of the umbrae over the solar cycle reported by \inlinecite{pevtsov2011}.  The size of the largest sunspots increases and, then, decreases as a function of the solar cycle as shown by \inlinecite{watson2011}, which in turn induces a solar cycle variation in the average umbral field strength inferred from the largest sunspots only. 

Due to the nonlinear relationship between magnetic field strength and umbral area, the magnetic field PDF may show small variations as a function of of the solar cycle evolution simply due to the change in the cycle amplitude.  This fact, coupled with the fact that the distribution of umbral areas on the Sun at a particular time may deviate significantly from the average umbral area PDF, can lead to variations in the mean umbral magnetic field strength as a function of time.  Such changes would not necessarily imply a significant deviation the solar dynamo activity, as suggested by \inlinecite{livingston2012}.  Still, in opposition to \inlinecite{livingston2012}, we do not find evidence for any long-term decrease in the mean magnetic field strength for the analyzed period. 

Furthermore, our determined magnetic field PDF does not support the normal distribution reported by \inlinecite{livingston2012}.  We note that our work has a considerably larger sample of umbrae.  According to \inlinecite{livingston2006} and \inlinecite{livingston2012}, the \textit{McMath-Pierce Telescope} 1565 nm umbral magnetic field data set includes 900 sunspots, measured across the entire solar disk between 1998 and 2005, and a total of 300 observing days between 1998 and 2011, amounting to an average sampling rate of $6\%$.  Meanwhile, our \textit{Hinode}/SP data set includes a total of 7530 umbrae between 2006 and 2012 during a time of low solar cycle ampltiude;  596 of these 7530 umbrae are unique, as they are measured close to solar meridian.  \inlinecite{livingston2012} do not record size information on the measured sunspots that can be compared with the previous observations of the umbral size spectrum.  Their observations of a observations of a core magnetic field distribution, whose shape remains unchanged while it mean decreases in time is inconsistent with our \textit{Hinode}/SP measurements.  While we assert that the magnetic field PDF can vary over the solar cycle, this variation is restricted primarily to the upper tail of the distribution.  Due to the relative sparseness of the measured spots between 1998 and 2002 in the \textit{McMath-Pierce Telescope} data set, the reported yearly change in the arithmetic mean is more suggestive of a sampling bias.  

To conclude, the collective study of simple measures of umbral properties provides constraints on the processes of continuum sunspot formation, and the connection of these measures with the bulk activity of the solar dynamo.  Understanding the variation of these mean properties over time is complicated because of the inherently nonlinear growth and evolution of individual sunspots.   Furthermore, we must be careful to not assume that the properties of large umbrae to hold for smaller dark features, as in the example of the radial field approximation for dark umbral cores.  Yet, viewing sunspot umbral cores as a continuum of sunspot formation seems to be valid as the relationship between magnetic field and temperature provide tight coherent relationships, albeit nonlinear and reflective of involved radiative MHD processes, that might be helpful in comparison with advanced modeling efforts such as \inlinecite{rempel2009}.


\begin{acks}

TAS is a research associate at the NSO. The NSO is operated by the Association of Universities for Research in Astronomy, Inc. (AURA), under cooperative agreement with the National Science Foundation.  Thanks are extended to Sanjay Gosain and Fraser Watson for helpful comments, as well as to the referee for constructive input.  Hinode is a Japanese mission developed and launched by ISAS/JAXA, collaborating with NAOJ as a domestic partner, NASA and STFC (UK) as international partners. Scientific operation of the Hinode mission is conducted by the Hinode science team organized at ISAS/JAXA. This team mainly consists of scientists from institutes in the partner countries. Support for the post-launch operation is provided by JAXA and NAOJ (Japan), STFC (U.K.), NASA, ESA, and NSC (Norway).  

\end{acks}


\bibliographystyle{spr-mp-sola-cnd} 
\bibliography{schad_bibliography} 

\end{article} 
\end{document}